\documentclass[onecolumn,preprint,noshowkeys,noshowpacs]{revtex4}%
\usepackage{amssymb}
\usepackage{amsmath}
\usepackage{amsfonts}
\usepackage{graphicx}%
\providecommand{\U}[1]{\protect\rule{.1in}{.1in}}

\begin{document}
\title{Thermal entanglement between non-nearest-neighbor spins on fractal lattices}
\author{Yu-Liang Xu, Lu-Shun Wang}
\author{Xiang-Mu Kong}
\thanks{Corresponding author}
\email{kongxm@mail.qfnu.edu.cn (X.-M. Kong)}
\affiliation{Shandong Provincial Key Laboratory of Laser Polarization and Information
Technology, Department of physics, Qufu Normal university, Qufu 273165, China}
\date{\today}

\begin{abstract}
We investigate thermal entanglement between two non-nearest-neighbor
sites in ferromagnetic Heisenberg chain and on fractal lattices by
means of the decimation renormalization-group (RG) method. It is
found that the entanglement decreases with increasing temperature
and it disappears beyond a critical value $T_{c}$. Thermal
entanglement at a certain temperature first increases with the
increase of the anisotropy parameter $\Delta$ and then decreases
sharply to zero when $\Delta$ is close to the isotropic point. We\
also show how the entanglement evolves as the size of the system $L$
becomes large via the RG method. As $L$ increases, for the spin
chain and Koch curve the entanglement between two terminal spins is
fragile and vanishes when $L\geq17$, but for two kinds of
diamond-type hierarchical (DH) lattices the entanglement is rather
robust and can exist even when $L$ becomes very large. Our result
indicates that the special fractal structure can affect the change
of entanglement with system size.

\end{abstract}
\maketitle

\section{Introduction}

An essential difference between quantum and classical physics is the
possible existence of nonlocal correlation in quantum system which
is called the entanglement \cite{1}. Recently, the quantum
entanglement has been recognized as an crucial resource in various
fields of quantum information such as quantum communication and
computation \cite{2,3,4,5,6,7}. Since the entanglement is fragile
and sensitive to many environment factors, many efforts are devoted
to studying stable entanglement for realistic system in finite
temperature. Thus, thermal entanglement has naturally received much
attention by its advantage of stability and requiring neither
measurement nor controlled switching of interactions in the
preparing process. For spin systems can be used for gate operation
in\ quantum computer thermal entanglement on solid spin systems have
been widely studied, for example, the Heisenberg spin chain both in
the absence \cite{absence field}\ and presence \cite{presence field
1, presence field 2}\ of an external magnetic field , spin rings
\cite{spin rings} and spin clusters \cite{spin clusters 1, spin
clusters 2}. However, most of these works only focused on thermal
entanglement between nearest, next-nearest or next-to-next-nearest
neighbor spins \cite{14,15,16,17,18,19}. This motivate us to propose
two questions: (i) Can thermal entanglement exist between distant
non-nearest-neighbor sites in spin system? (ii) How does thermal
entanglement evolve as system size grows? But it is very difficult
to obtain exact results on entanglement in spin systems on arbitrary
lattices especially fractal lattices, since this usually requires
the expression of the partition function which is too complicated to
solve when the system size becomes very large.

In recent years, the entanglement at zero temperature in the large
size system has been studied by adopting the renormalization-group
(RG) method. In 2002, A. Osterloh \textit{et al} first introduced
the density matrix renormalization-group (DMRG) approach to study
the entanglement close to the quantum phase transition (QPT)
\cite{Nature} and reveal a profound difference between classical
correlations and the non-local quantum correlation. Further,by
applying the quantum renormalization-group (QRG) approach, M.
Kargarian \textit{et} \textit{al } investigated the entanglement in
the anisotropic Heisenberg model \cite{XXZ 1, XXZ 2} and discussed
the nonanalytic behaviors and the scaling close to the quantum
critical point of the system. Recently We have calculated the
block-block entanglement in the \textit{XY} model without and with
staggered Dzyaloshinskii-Moriya (DM) interaction by using this QRG
method and have found the DM interaction can enhance the
entanglement and influence the QPT of the system \cite{Fu-Wu Ma,fu
wu ma}.

Inspired by above idea, we apply the real-space
renormalization-group (RSRG) approach which is developed in the
Refs. \cite{24,25,26,27,28,29,30,31,32,33,34} to study the thermal
entanglement between two end sites in the spin chain, Koch curve and
on the diamond-type hierarchical (DH) lattices and analyze the
influence of the temperature, the anisotropy parameter and the
system size on the entanglement.

\section{Model and method}

The effective Hamiltonian of the spin-$1/2$ anisotropic
ferromagnetic Heisenberg spin chain with $L$ sites is
\begin{equation}
-\beta H=\sum_{i}^{L}K\left[  \left(  1-\Delta\right)  \left(  \sigma_{i}%
^{x}\sigma_{i+1}^{x}+\sigma_{i}^{y}\sigma_{i+1}^{y}\right)  +\sigma_{i}%
^{z}\sigma_{i+1}^{z}\right]  ,\label{1}%
\end{equation}
where $\sigma_{i}^{\alpha}$ $\left(  \alpha=x,y,z\right)  $ denote the Pauli
operators at site $i$. $K=\beta J=J/k_{B}T$, $J>0$ is the exchange coupling
parameter, and $k_{B}$ is the Boltzmann constant. For simplicity, we assume
that $k_{B}=1$ and $J=1$. The sum is over all the nearest-neighbor spin pairs
and $\Delta\in(-\infty,1]$ is the anisotropy parameter. For $\Delta=0$ and
$\Delta=1$, the isotropic Heisenberg (\textit{XXX}) and Ising model are
obtained, respectively. The state of the above system in thermal equilibrium
can be described by the density operator $\rho=Z^{-1}\left(  e^{-\beta
H}\right)  $, where $Z=$Tr$\left(  e^{-\beta H}\right)  $ is the partition function.

The entanglement of two-qubit system in the thermal state $\rho_{_{12}}$ can
be calculated by the negativity \cite{Negativity} which is based on the
partial transpose method \cite{Partial transpose}. The negativity $N$ is
defined as
\begin{equation}
N\left(  \rho_{_{12}}\right)  =2\sum_{i}\left\vert \mu_{i}\right\vert ,
\label{2}%
\end{equation}
where $\mu_{i}$ is the negative eigenvalue of $\rho_{12}^{T_{1}}$,
and $T_{1}$ denotes the partial transpose with respect to the first
subsystem. According this definition one can easily obtain thermal
entanglement $N\left( K^{\prime},\Delta^{\prime }\right)  $ of the
two-spin chain with the Hamiltonian $H_{12}^{\prime}\left(
K^{\prime},\Delta^{\prime}\right) $ (such as Fig 1. (a) $n=0$
shown). However, when the size of system becomes large, the density
matrix is difficult or impossible to gain. The entanglement of two
terminal spins on this system can not be directly worked out.

We apply the decimation RSRG method to solve the above problem. This
decimation RSRG method \cite{26,28,37} has proved to be successful
in spin chain and especially the fractal lattices.  For the spin
chain, this decimation procedure is illustrated in Fig. 1 (a).
Simply, the generator is taken out in the infinite system. The
generator with the Hamiltonian $H_{132}\left( K,\Delta\right)  $ is
renormalized into the new two-site chain with the Hamiltonian
$H_{12}^{\prime}\left(  K^{\prime},\Delta^{\prime}\right)  $ by
integrating the internal site $3$ with the partition function being
preserved.
This transformation can be described as%
\begin{equation}
\exp\left(  H_{12}^{\prime}\right)  =\text{Tr}_{3}\exp\left(  H_{132}\right)
. \label{3}%
\end{equation}

We can obtain the recurrence relation between the original parameters $\left(
K,\Delta\right)  $ and the new parameters $\left(  K^{\prime},\Delta^{\prime
}\right)  $ by solving the trace Tr$_{3}$ with the method developed in Refs
\cite{28,30}. Combining this relation $K^{\prime}=g\left(  K,\Delta\right)  ,$
$\Delta^{\prime}=h\left(  K,\Delta\right)  $ and the negativity $N\left(
K^{\prime},\Delta^{\prime}\right)  $ of two-site system, the\ entanglement
between two terminal spins in three-qubit chain can be obtained as follow%
\begin{equation}
N\left(  K^{\prime},\Delta^{\prime}\right)  =N\left(  g\left(  K,\Delta
\right)  ,\text{ }h\left(  K,\Delta\right)  \right)  . \label{4}%
\end{equation}

The entanglement between two distant terminal spins in the Heisenberg chain
can be calculated after many iterations of the recurrence relation. We take
use of the same method to study thermal entanglement between two terminal
sites in Koch curve and on the DH lattices as shown in Fig. 1 (b) (c) and (d).
The analytical expression about the negativity in Eq. (\ref{4}) is difficult
to obtain, we will show some numerical results.

\section{Heisenberg spin chain}

We first study how the entanglement between terminal sites in spin
chain with different number of sites $L$ varies with temperature $T$
at $\Delta=-0.2$ (shown in Fig. 2). For different cases of $L$, the
results have the similar feature that the entanglement decreases
monotonically with increasing temperature and it vanishes beyond the
critical temperature $T_{c}$. At $T=0$, the system is in the
entangled ground state. As temperature increases, the entanglement
decreases due to the mixture of the unentangled excited state with
the ground state. At $T=T_{c}$, the system is governed by the
unentangled excited state completely, and therefore the entanglement
vanishes. Comparing the entanglement of terminal sites in the
different system, it is found that the entanglement of the ground
state decreases sharply with increasing $L$. Different from the
maximally entangled Bell ground state for $L=2$ system, the ground
state for $L>2$ system becomes a degenerate and related to $\Delta$
state which cause the decrease of entanglement. The energy gap
between the ground and the unentangled excited state increases with
little range but the thermal energy of system increases with large
range when $L$ increases. For $L>2$, the system can easily overcome
the gap and enter the unentangled state. That leads to the decrease
of $T_{c}$. This phenomenon reflects that thermal fluctuation of
internal sites may suppress quantum effect.

The influence of the anisotropic parameter $\Delta$ on the
entanglement between two terminal sites at a fixed temperature
$T=0.01$ is plotted in Fig. 3. As can be seen, for $L=2$ case, the
system firstly keeps in the ground state with the entanglement
$N=1$. Then, as $\Delta$ approaches the isotropic point $\Delta=0$,
the energy gap between the ground and the unentangled excited states
becomes so small that the system can jump to the unentangled excited
state. Therefore there exists a sharp decease for entanglement and
the entanglement vanishes when $\Delta$ is close to zero. This
result also accords with that the entanglement can not exist in an
isotropic Heisenberg ferromagnetic chain in Ref \cite{Nielsen}. For
$L\neq2$ cases, it is found that the entanglement increases firstly
with increasing $\Delta$ because the ground state is related to
$\Delta$ and will change with $\Delta$. All entanglement jump down
to zero when $\Delta$ reaches to zero. From above results, we can
see that the entanglement is fragile and when $L\geq17$ the
entanglement does not exist whatever the temperature and the
anisotropic\ parameter are.

\section{Fractal lattices}

The properties of phase transition on different fractal lattices have been
studied by the RSRG method, and the entanglement on these self-similar
lattices remains to be explored. We turn to the study of the entanglement
between end sites in Koch curves with non-integer fractal dimension $d_{f}%
=\ln4/\ln3$ and plot the numerical results of negativity versus $T$ and
$\Delta$ for different $L$ in Fig. 4. Compared with the entanglement in the
spin chain ($L=5$), it has similar properties that the entanglement decreases
with $T$ and the maximal of entanglement and $T_{c}$ are approximately equal.
But the entanglement variation versus $\Delta$ is very different from that in
spin chain, i.e., the range that the entanglement can exist is smaller, the
maximal of entanglement is lager. The entanglement decreases quickly as $L$
becomes large and there exists no entanglement any longer when $L\geq17$. We
can deduce this result from the similar Hamiltonian and open boundary
conditions of these two systems.

Now we consider two kinds of DH lattices with fractal dimensions $d_{f}=2$
(lattice A, for simplicity) and $d_{f}=\ln5/\ln2$ (lattice B, for simplicity).
The RG transformation on these DH lattices respectively have been shown in
Fig. 1 (c) and Fig. 1 (d). We first discuss the dependence of the entanglement
between terminal sites on $T$ with $\Delta=-0.2.$ In Fig. 5, one can find some
similar behaviors that the entanglement is the maximal value at $T=0$, and the
entanglement decreases with increasing temperature and vanishes beyond the
critical temperature. However, some different phenomenons are also observed
that the entanglement between end sites on these two lattices decrease more
slowly with increasing $L$ and it still exists even though the system size
becomes very large ($L=1564$). For the case of on lattice B, as $L$ increases,
the entanglement at zero temperature decreases but the corresponding $T_{c}%
$\ increases. It is obvious that the entanglement for different\ $L$ crosses
at $T\approx0.49$. This result indicates that these two fractal lattices have
special energy level structure. The energy gap between the entangled ground
state and the unentangled excited state is so large that the system can jump
to the unentangled excited state only at higher $T$.

The variation of the entanglement between end sites on DH lattices versus
$\Delta$ at $T=0.01$ is also discussed. As can be seen in Fig. 6 (a) for the
lattice A, the entanglement firstly increases as $\Delta$ increases, and then
it quickly decays to zero when $\Delta$ reaches the isotropic point. The
entanglement decrease very slowly when $L$ becomes very large and there exist
no cross point when $\Delta$ reaches zero. For the case of lattice B, Fig. 6
(b) shows that the entanglement also exhibits stable and it changes very
little when $\Delta$ is not very close to zero. The entanglement mainly
remains robust with the increase of $L$. In this graph, we also observe that a
"entanglement crossing" occurs at $\Delta\approx-0.045$.\ At a fixed
temperature, the thermal excited energy of the system is determined. Only when
$\Delta$ is very close to zero, the energy gap between the entangled ground
state and the unentangled excited state can become so small that the system
can enters the unentangled state. It also indicates that the different fractal
structure can influence the entanglement by changing the energy level
structure of system.

\section{Conclusion}

We have investigated thermal entanglement between two end spins in Heisenberg
chain, Koch curve and on two kinds of DH lattices with $d_{f}=2$ and
$d_{f}=2.32$ by the decimation RG method. The effect of the temperature and
the anisotropy parameter on thermal entanglement is discussed. It is found
that the symmetry of system and the thermal fluctuation can suppress or
promote the quantum effect at different conditions. We also have noticed that
the entanglement on some special lattices may exhibit different property when
the system size $L$ becomes large. The entanglement on two kinds of DH
lattices is quite robust and it can survive even though $L$ becomes very large
in contrast to that in spin chain. The phenomenon of the "entanglement
crossing" indicates that the special fractal structure does influence on the entanglement.

\begin{acknowledgments}
This work is supported by the National Natural Science foundation of China
under Grant No. 10775088, the Shandong Natural Science foundation under Grant
No. Y2006A05, and the Science foundation of Qufu Normal University. The
authors would like to thank Hong-Bing Li, Yin-Fang Li, and Cong-Fei Du for
fruitful discussions and useful comments.
\end{acknowledgments}

\newpage\ \ \ Figure captions:

Fig. 1. The procedure of the RG transformation. From (a) to (d), it shows the
transformation of one-dimensional spin chain, Koch curve, the diamond-type
hierarchical lattice with fractal dimension $d_{f}=2$ and $d_{f}=\ln5/\ln2$.

Fig. 2. The entanglement between two end sites in Heisenberg ferromagnetic
chain versus temperature at $\Delta=-0.2$ for different number of sites $L$
(from top to bottom, $L=2$, $3$, $5$ and $9$).

Fig. 3. The entanglement between two end sites in Heisenberg ferromagnetic
chain as anisotropy parameter $\Delta$ at $T=0.01$ for different number of
sites $L$ (from top to bottom, $L=2$, $3$, $5$ and $9$).

Fig. 4. The variation of the entanglement between end sites in Koch curve: (a)
the entanglement versus temperature $T$ at $\Delta=-0.2$. (b) the entanglement
versus $\Delta$ at $T=0.01$.

Fig. 5. Upper panel: the negativity of two end sites on the DH lattice with
$d_{f}=2$ versus $T$ at $\Delta=-0.2$. Lower panel: the negativity of two end
sites on the DH lattice with $d_{f}=\ln5/\ln2$ versus $T$ at $\Delta=-0.2$.
The negativity for different value of $L$ has a cross point at $T\approx0.49$.

Fig. 6. Upper panel: the evolution of the entanglement between terminal sites
on DH lattice with $d_{f}=2$ as $\Delta$ increases at $T=0.01$ for different
value of $L.$ Lower panel: the entanglement between terminal sites on DH
lattice with $d_{f}=\ln5/\ln2$ versus $\Delta$ at $T=0.01$ for different value
of $L$. The entanglement for different $L$ has a cross point at $\Delta
\approx-0.045$.


\begin{thebibliography}{99}                                                                                               %


\bibitem {1}J. S. Bell, Physics (Lang Island city), \textbf{1}, 195 (1964).

\bibitem {2}Artur K. Ekert, Phys. Rev. Lett. \textbf{67}, 661 (1991).

\bibitem {3}C.H. Bennett and S.J. Wiesner, Phys. Rev. Lett. \textbf{69,} 2881 (1992).

\bibitem {4}C.H. Bennett \textit{et al}, Phys. Rev. Lett. \textbf{70}, 1895 (1993).

\bibitem {5}B. E. Kane, Nature (Londond) \textbf{393}, 133 (1998).

\bibitem {6}C. H. Bennett and D. P. Divincenzo, Nature (Londond) \textbf{404},
247 (2000d).

\bibitem {7}M. A. Nielson and I. L. Chuang, \textit{Quantum Computation and
Quantum Information} (Cambridge University Press, Cambridge, U.K., 2000).

\bibitem {absence field}W. K. Wootters, Phys. Rev. A \textbf{63}, 052302 (2001).

\bibitem {presence field 1}M. C. Arnesen, S. Bose, V. Vedral, and S. Bose,
Phys. Rev. Lett. \textbf{87}, 017901 (2001).

\bibitem {presence field 2}X. Wang, Phys. Rev. A \textbf{64}, 012313 (2001).

\bibitem {spin rings}X. Wang, Phys. Rev. A \textbf{66}, 034302 (2002).

\bibitem {spin clusters 1}Indrani Bose and Amit Tribedi, Phys. Rev. A
\textbf{72}, 022314 (2005).

\bibitem {spin clusters 2}A. M. Souza, M. S. Reis, D. O. Soares-Pinto, I. S.
Oliveira and R. S. Sarthour, Phys. Rev. B \textbf{77}, 104402 (2008).

\bibitem {14}K. Audenaert, J. Eisert, and M. B. Plenio R. F. Werner Phys. Rev.
A \textbf{66}, 042327 (2002).

\bibitem {15}L. Zhou, H. S. Song, Y. Q. Guo, and C. Li, Phys. Rev. A
\textbf{68}, 024301 (2003).

\bibitem {16}Guo-Feng Zhang, and Shu-Shen Li, Phys. Rev. A \textbf{72}, 034302 (2005).

\bibitem {17}Zhao-Yu Sun, Kai-Lun Yao, Wei Yao, De-Hua Zhang, and Zu-Li Liu,
Phys. Rev. B \textbf{77}, 014416 (2008).

\bibitem {18}Fardin Kheirandish, S. Javad Akhtarshenas, and Hamidreza
Mohammadi Phys. Rev. A \textbf{77}, 042309 (2008).

\bibitem {19}A. M. Souza, M. S. Reis, D. O. Soares-Pinto, I. S. Oliveira, and
R. S. Sarthour, Phys. Rev. B \textbf{77}, 104402 (2008).

\bibitem {Nature}A. Osterloh, L. Amico, G. Falci, and R. Fazio, Nature
\textbf{416}, 608 (2002).

\bibitem {XXZ 1}M. Kargarian, R. Jafari, and A. Langari, Phys. Rev. A
\textbf{76}, 060304 (2007).

\bibitem {XXZ 2}M. Kargarian, R. Jafari and A. Langari, Phys. Rev. A
\textbf{77}, 032346 (2008)

\bibitem {Fu-Wu Ma}Fu-Wu Ma, Sheng-Xin Liu and Xiang-Mu Kong, Phys. Rev. A
\textbf{83}, 062309 (2011)

\bibitem {fu wu ma}Fu-Wu Ma, Sheng-Xin Liu and Xiang-Mu Kong, Phys. Rev. A
\textbf{84}, 042302 (2011).

\bibitem {24}A. P. Young and R. B. Stinchcombe, J. Phys. C: Solid State Phys
\textbf{9}, 4419 (1976).

\bibitem {25}R. Jullien, J. Fields and S. Doniach, Phys. Rev. Lett.
\textbf{38}, 1500 (1977).

\bibitem {26}M. Suzuki and H. Takano, Phys. Lett. A \textbf{69}, 426 (1979).

\bibitem {27}H. Takano and M. Suzuki, J. Stat. Phys. \textbf{26}, 635 (1981).

\bibitem {28}Anibal O. Caride, Constantino Tsallis, and Susana I. Zanette,
Phys. Rev. Lett. \textbf{51}, 145 (1983).

\bibitem {29}A. M. Mariz, R. M. Zorzenon Dos Santos, C. Tsallis and R. R. Dos
Santos, Phys. Lett. A \textbf{108}, 95 (1985).

\bibitem {30}A. M. Mariz, C. Tsallis and A. O. Caride, J. Phys. C \textbf{18}
4198 (1985).

\bibitem {31}S. R. White and R. M. Noack, Phys. Rev. Lett. \textbf{68}, 3487 (1992).

\bibitem {32}J. Ricardo de Sousa, Phys. Lett. A \textbf{216}, 321 (1996).

\bibitem {33}N. S. Branco, J. Ricardo de Sousa, Phys. Rev. B \textbf{62}, 5742 (2000).

\bibitem {34}J. Ricardo de Sousa, N. S. Branco, B. Boechat and Claudette
Cordeiro, Physica A \textbf{328}, 167 (2003).

\bibitem {Negativity}G. Vidal, and R. F. Werner, Phys. Rev. A \textbf{65},
032314 (2002).

\bibitem {Partial transpose}A. Peres, Phys. Rev. Lett. \textbf{77}, 1413
(1996); M. Horodecki, P. Horodecki and R. Horodecki, Phys. Lett. A
\textbf{223}, 1 (1996).

\bibitem {37}Andre M. C. de Souza, Phys. Rev. B \textbf{48}, 3744 (1993).

\bibitem {Nielsen}M. A. Nielsen, Ph. D. thesis, University of Mexico, 1198,
e-print arXiv: quant-ph/0011036.
\end{thebibliography}
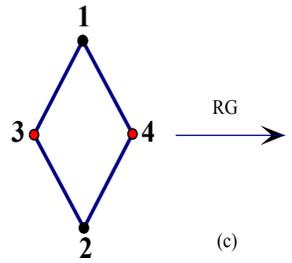
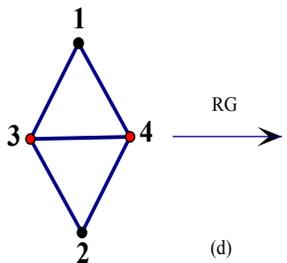
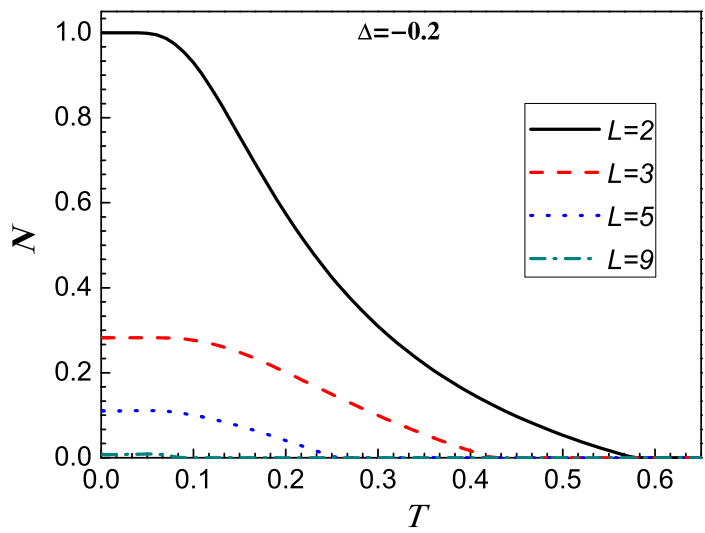
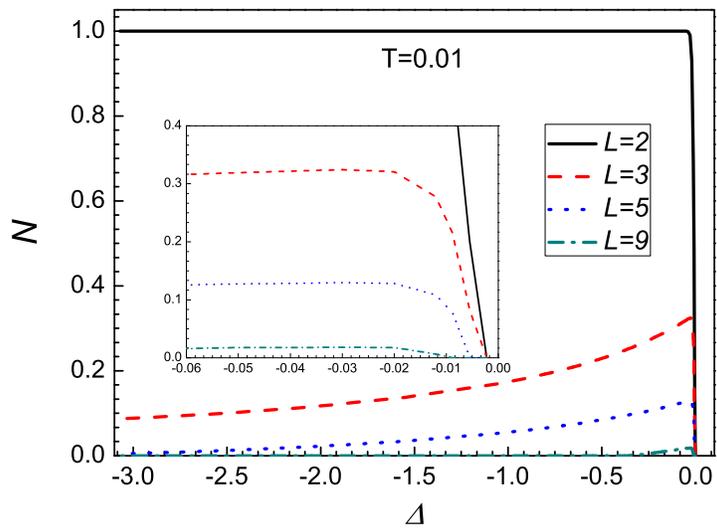
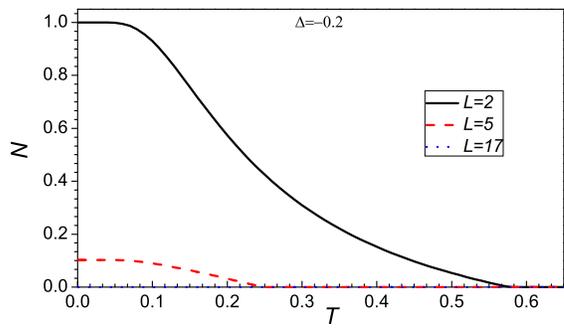
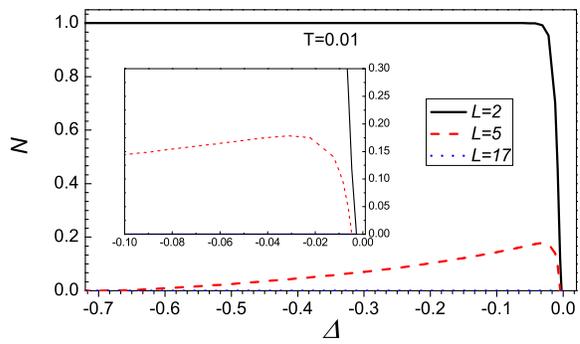
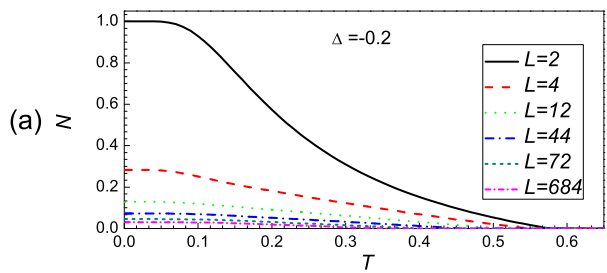
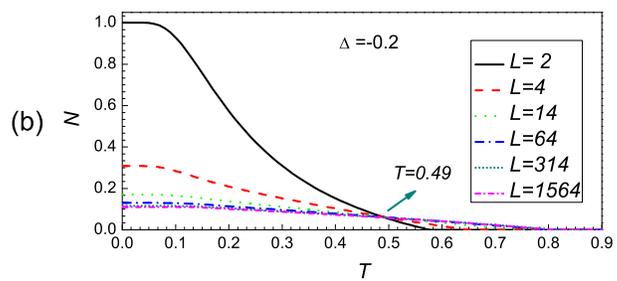
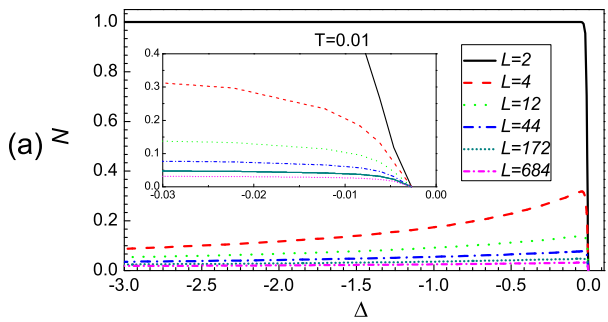
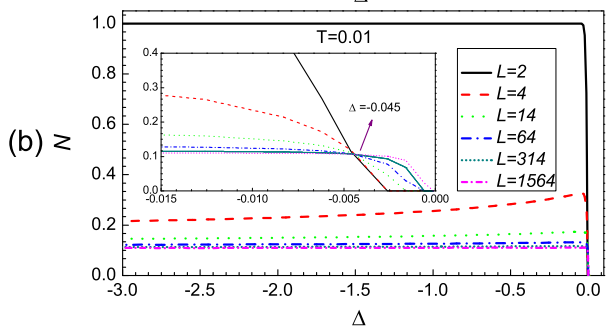
\end{document}